# User Profiling for Recommendation System

Sumitkumar Kanoje
Dept. of Information Technology
MIT Pune, India
sumitkanoje@gmail.com

Sheetal Girase
Dept. of Information Technology
MIT Pune, India
girase.sheetal@gmail.com

Debajyoti Mukhopadhyay
Dept. of Information Technology
MIT Pune, India
debajyoti.mukhopadhyay@gmail.com

*Abstract*—Recommendation system is a type of information filtering systems that recommend various objects from a vast variety and quantity of items which are of the user interest. This results in guiding an individual in personalized way to interesting or useful objects in a large space of possible options. Such systems also help many businesses to achieve more profits to sustain in their filed against their rivals. But looking at the amount of information which a business holds it becomes difficult to identify the items of user interest.

Therefore personalization or user profiling is one of the challenging tasks that give access to user relevant information which can be used in solving the difficult task of classification and ranking items according to an individual's interest. Profiling can be done in various ways such as supervised or unsupervised, individual or group profiling, distributive or and non-distributive profiling. Our focus in this paper will be on the dataset which we will use, we identify some interesting facts by using Weka Tool that can be used for recommending the items from dataset .Our aim is to present a novel technique to achieve user profiling in recommendation system.

## I. INTRODUCTION

A User Profile is a set of features and/or patterns used to concisely describe the user. User Profiling is a process especially critical for e-business systems to capture online users' characteristics, know online users, provide customized products and services, and therefore improve user satisfactions. User profiling techniques have widely been applied in various e-business applications, e.g. online customer segmentation, web user identification, adaptive web site, fraud/intrusion detection, personalization, recommendation, e-market analysis, as well as personalized information retrieval and filtering.

Profiling of a Web user is the process of obtaining values of different properties that constitute the user model. Considerable efforts have been made to mine the user's interests from his/her historical data. A typical way for representing the user's interests is to create a list of relevant keywords. However, such a profile is insufficient for modeling and understanding users' behavior. A complete user profile (including one's education, experience, and contact information) is very important for providing high-quality Web services. For example, with a well-organized user profile base, online advertising can be more targeted based on not only on a user's interests but also on his/her current position.

We are developing a system which will recommend various universities to the user who is using our system. The system is a combination of various approaches for providing relevant universities to user. This involves much of task of information retrieval from various sources.

## II. MOTIVATION

A static user profiling approach includes explicitly filling information from the user through the use of online forms and surveys. This technique is a simplest and easiest way of creating profiles from the information entered by the user. But most of the users are not interested to reveal their information to anyone as they are concerned about their privacy or due to the tediousness of form filling process. Also it has been observed there is a risk of user entering fraudulent or wrong information due to such tediousness of this process. Hence the accuracy of using this type of profiling degrades as user might give false information about himself.

So there is need of a system which will automatically try to retrieve the users' information from several other sources. This is where Social Information Discovery and User Profiling play an important role. Peoples tend to take their friends' opinion before making their own choice. So an implicit user profiling through social discovery will help resolving long-tail problem on user profiling in this report is focused on factual extracting, integrating constructing and visualizing user profiles.

As users are not interested in disclosing their information directly there is a big challenge of knowing the user. The task of user profiling gets tougher without having at least some information about the user. Now day's data is available everywhere but gathering those data and finding something interesting out of it is a big challenge. An innovation is necessary in extracting that information through various ways from the user.

## III. RELATED WORK

Recommender systems can be considered as the direct beneficiary of user profiling, user profiling is an important part of the recommender systems since earlier times. But nowadays user profiling is becoming common in many of the applications like Search Personalization [15], Adaptive Websites, Adaptive Web stores and Customer Relationship Management systems. Some of the case studies of such applications are given in this section.

User profiling for recommendation of research papers is an application where much of the work has been done. The system developed by Tang [2] has divided the task of user profiling into three subtasks viz. profile extraction, integration and interest discovery. In a similar approach by Stuart Middleton [4] used one extra step viz. profile visualization to represent a profile generated by the system which used ontological approach.

e-Tourism based website [13] is another application which can be benefited by User Profiling. This system was able to deliver personalized information based on the location of the user. As the tourism business is totally dependent on the demographic information like location the system was able to provide recommendations of nearest tourist spots to a new user based on his location.

Energy management is a very important task of nowadays. There are many big enterprises which are facing the challenge of efficient and optimized energy management. The smart energy management system developed by [12] have proved to be efficient this task. Here they used user profiling and micro accounting for smart energy management.

Finding a job is one of the tedious jobs everyone has to do in his lifetime. So developing a system which will automatically recommend jobs to a user as per his qualification and experiences is one the great idea which this author has succeeded to implement. This system is called CASPER [14] (Case-Based Profiling for Electronic Recruitment). The system takes into account users profiling information and recommends suitable jobs to every individual.

## IV. PROBLEM STATEMENT

When humans come across making choice from large information they find it difficult to obtain the most relevant information that is hidden in the deluge of information. When there is mass of content available with us, important questions is raised over its effective use.

Recommender systems provide advice to users about items they might be interested in. Recommendations made by such systems can help users navigate through large information spaces of product descriptions, news articles or other items. Recommending such items according to user interest involves processing through these large digitized information spaces; if this information is already being profiled properly it be easier for recommendation system to recommend them to user. User profiling comes into picture in this scenario.

User profiling has wide applications such as personalization, intrusion detection, and online customer analysis in e-business environments. Profiling of a Web user is the process of obtaining values of different properties that constitute the user model. User profiling is typically either knowledge-based (already known/factual) or behavior-based. A typical user profiling system is aimed at finding, extracting, and fusing the keyboard based user profile from the Web.

The problem statement revolves around profiling. We will extract the basic information about a university, location information, size information, ranking information etc. Also we will be profile users in the system implicitly by extracting his information from social networks. This is done by finding, extracting, integrating and profiling the keyword based information of the researcher from web and then visualizing this profile on a web page.

Basic goal is to create a profile for each University, which contains basic information e.g., name, accreditation, contact information (e.g., address, email, and telephone number), course information (e.g., degree courses, post graduate courses, and Phd courses), research interests. For each user, some of the profile information can be extracted from his/her social networking websites introducing him/her; and the other information (e.g., research interests) can be mined from the collected information.

Now we implement the techniques for various concepts like implicit user profiling. These concepts are the part of profile extraction, integration and user interest discovery. A search facility will also be implemented for retrieval of particular particular profile.

## V. PROPOSED METHODOLOGY

This work is organized as the task of user profiling to achieve following objectives
- Profile Extraction for each University
- Profiling of every User
- Profile Integration on extracted information
- User Interest Discovery from the available information

We propose a novel approach to solve the user profiling problem. System architecture in the next section shows the overview of this approach. There are mainly three components: profile extraction and integration, and user interest analysis. The first component targets extracting the profiles, second integrating profile information from the Web; and the third targets analyzing users' interests.

### A. Profile extraction

Profile extraction is nothing but extracting the useful information about a user from different sources. In our application we have to profile both university and the user. The university so we have modeled our system in two basic profile creation steps viz. University Profile & User Profile.

#### 1) University Profile

For university profiling identifying different attributes is necessary. Fig represents the university schema that will be prepared after extracting the university profile.

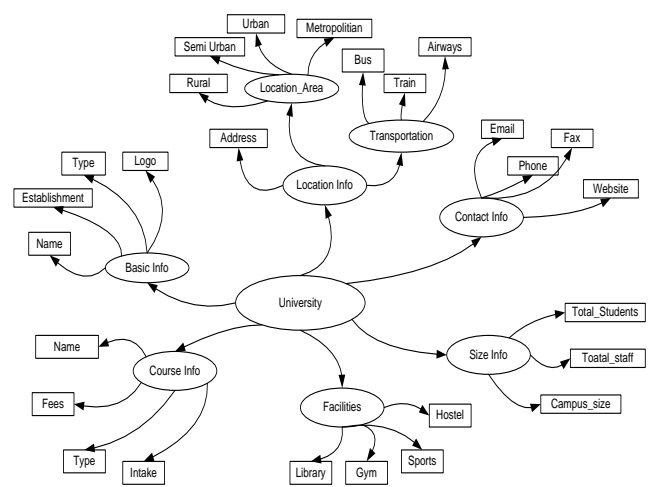

Figure 1. University Schema

In pre-processing, (a) we segment the text into tokens and (b) we assign possible tags to each token. The tokens form the basic units and the pages form the sequences of units or a tree structure of units in the tagging problem. In tagging, given a sequence of units or a tree structure of units, we determine the most likely corresponding tags using a trained tagging model. Each tag corresponds to a property defined in Figure 2. In this system, we use a Tree-structure Conditional Random Fields (TCRF) [2] as the tagging model.

*2) User Profile*

User profile extraction is the process in which we get users information while the user registers into our system. For profiling user we identify different attributes of user.

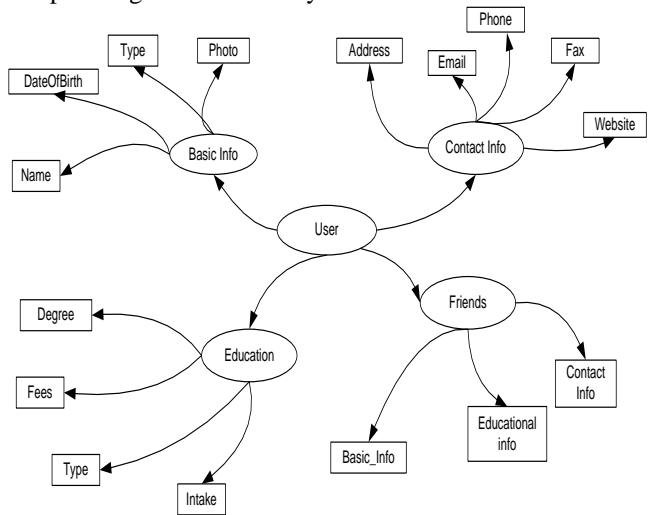

Figure 2: User Profile

To allow implicit profiling we do extract the users' details from different social network and profile the user.

*B. Profile Integration*

We crawled the university data from existing online data sources. For integrating the extracted data from several sources we use profile integration. The method inevitably has the name ambiguity problem.

For integrating the extracted from in previous phase we have to integrate this data from various sources in a single format which will be same for all the data. We create a dataset in this step which will be used for our further operations such as interest discovery. We remove various attributes with missing values also we refine the data as per our requirements for example we can form data value ranges instead of keeping the exact value. This helps our task of interest discovery.

*C. Interest Discovery*

After extracting and integrating the user profiles, we obtain a basic user profile which consists of a set of profile properties and a set of documents for each user. Now we perform user interest analysis based on the user profile and its associated papers.

According to the definition of user interest our goal is to discover the latent topic distribution associated with each user. For different applications, the available information to discover the latent topic distribution is also different. As in case of our system, available information includes university location, university control, no of students in university, university facilities and other related information about university. Also users' educational background, his location, his visited places etc. is present in his profile extracted in the previous steps.

Also to implicitly know the user interests we will be providing a search facility which will get the search terms of the users and these will be used for profiling the user. Now this information can be used to efficiently know user interests.

In the model, each user is associated with a multinomial distribution over attributes and each word token in a university profile and the interested universities are generated from a sampled topic.

## VI. SYSTEM ARCHITECTURE

Below figure gives an overview of our system architecture.

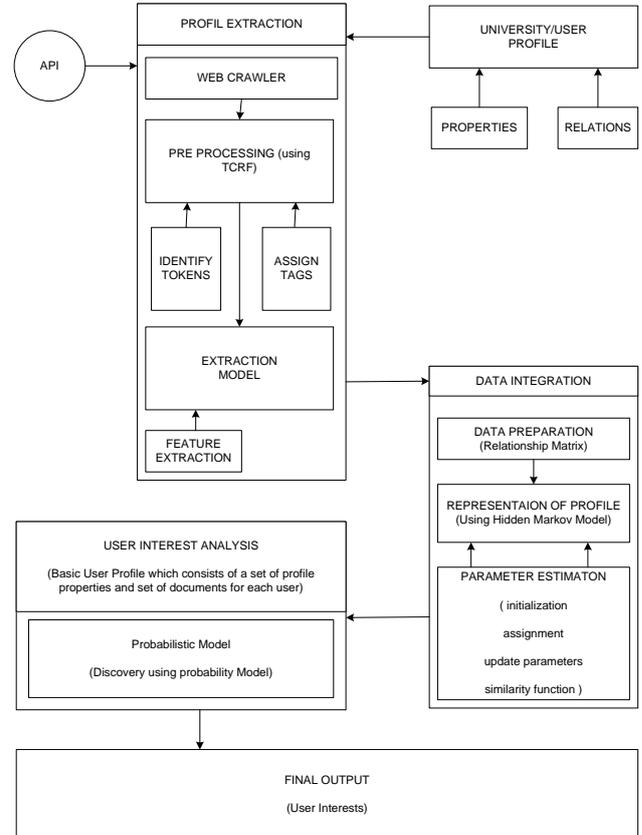

Figure 3: System Architecture

## VII. PROFILING OUTCOMES

For initial experiment we used university database provided by UCI Machine Learning Repository [17] which included total 285 instances of universities provided as .data file.

To gain knowledge out of it we decide to process this dataset in Weka Tool. As this dataset has several duplicates and multi-valued attributes which are not supported by Weka Tool we pre-processed this dataset to remove duplicate values from it. Also we used different attributes to move over the problem of multi-valued attributes. Finally we are done with a new dataset created in Weka Tool's standard .arff format. The arff dataset is shown below.

```
1  @relation Universities-v2
2
3  @attribute name string
4  @attribute state string
5  @attribute location {SUBURBAN,URBAN,SMALL-TOWN,SMALL-CITY}
6  @attribute control {PRIVATE,STATE}
7  @attribute no-of-students {5-,05-10,15-20,20+}
8  @attribute expenses {4-,04-07,07-10,10+}
9  @attribute percent-financial-aid numeric
10 @attribute number-of-applicants {01-10,04-07,07-10,17+,13-17,4-}
11 @attribute percent-admittance numeric
12 @attribute percent-enrolled numeric
13 @attribute academics {1,2,3,4,5}
14 @attribute social {1,2,3,4,5}
15 @attribute quality-of-life {1,2,3,4,5}
16 @attribute academic-emphasis-arts {YES,NO}
17 @attribute academic-emphasis-science {YES,NO}
18 @attribute academic-emphasis-commerce {YES,NO}
19 @attribute academic-emphasis-engg {YES,NO}
20 @attribute academic-emphasis-mangment {YES,NO}
21 @attribute academic-emphasis-education {YES,NO}
22 @attribute academic-emphasis-medical {YES,NO}
23
24 @data
25 ADELPHI,NEWYORK,?,PRIVATE,05-10,07-10,60,04-07,70,40,2,2,2,NO,YES,NO,NO,YES,NO,NO
26 ARIZONA-STATE,ARIZONA,?,STATE,15-20,04-07,50,17+,80,60,3,4,5,YES,NO,YES,YES,YES,NO,NO
27 BOSTON-COLLEGE,MASSACHUSETTS,SUBURBAN,PRIVATE,05-10,10+,60,01-10,50,40,4,5,3,YES,YES,YES,NO,NO,NO,NO
28 BOSTON-UNIVERSITY,MASSACHUSETTS,URBAN,PRIVATE,05-10,10+,60,13-17,60,40,4,4,3,YES,NO,NO,NO,YES,NO,NO
29 BROWN,RHODEISLAND,URBAN,PRIVATE,5-,10+,40,01-10,20,50,5,4,5,YES,YES,NO,NO,NO,NO
30 CAL-TECH,CALIFORNIA,SUBURBAN,PRIVATE,5-,10+,70,4-,15,90,5,1,3,NO,NO,NO,YES,NO,NO,NO
31 CARNEGIE-MELLON,PENNSYLVANIA,URBAN,PRIVATE,5-,10+,70,04-07,40,50,4,3,3,NO,NO,NO,YES,NO,NO
32 CASE-WESTERN,OHIO,URBAN,PRIVATE,5-,10+,65,4-,85,35,3,2,3,YES,YES,NO,YES,NO,YES
33 CCNY,NEWYORK,URBAN,STATE,05-10,4-,80,4-,80,60,3,2,2,YES,YES,NO,YES,NO,YES
```

Figure 4: Processed arff Dataset

After processing the dataset from Weka Tool we have generated the following output as shown in the below figure. Some interesting facts obtained can be seen in below figure.

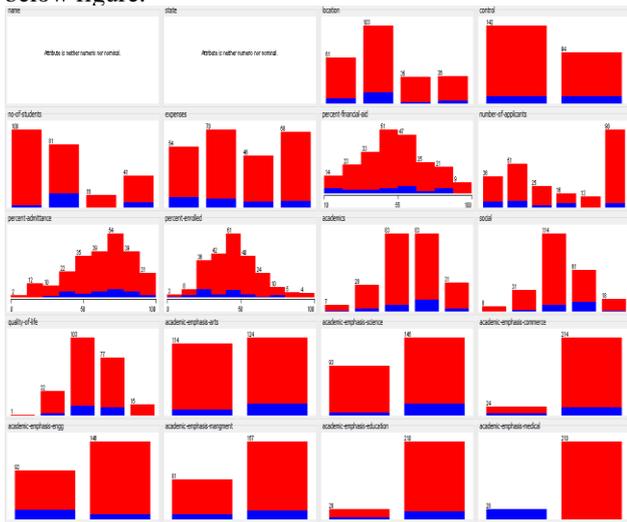

Figure 5: Results Generated from Weka Tool

From the obtained results we have grouped universities different classes based on their values. The outputs are shown below.

TABLE I. UNIVERSITY CLASSES BASED ON LOCATION

| Sr. No | University Classes | | |
|---|---|---|---|
| | Class | Count | % |
| 1 | SUBURBAN | 61 | 25.63 |
| 2 | URBAN | 103 | 43.27 |
| 3 | SMALL-TOWN | 35 | 14.70 |
| 4 | SMALL-CITY | 36 | 15.12 |

TABLE II. UNIVERSITY CLASSES BASED ON CONTROL

| Sr. No | University Classes | | |
|---|---|---|---|
| | Class | Count | % |
| 1 | PRIVATE | 61 | 25.63 |
| 2 | STATE | 103 | 43.27 |

TABLE III. UNIVERSITY CLASSES BASED ON No OF STUDENTS

| Sr. No | University Classes | | |
|---|---|---|---|
| | Class | Count | % |
| 1 | <5 | 100 | 42.01 |
| 2 | 5-10 | 81 | 34.03 |
| 3 | 10-20 | 16 | 6.72 |

TABLE IV. UNIVERSITY CLASSES BASED ON EXPENSES

| Sr. No | University Classes | | |
|---|---|---|---|
| | Class | Count | % |
| 1 | 0-4 | 54 | 22.68 |
| 2 | 4-7 | 70 | 29.41 |
| 3 | 7-10 | 46 | 19.32 |
| 4 | 10+ | 68 | 28.57 |

TABLE V. UNIVERSITY CLASSES BASED ON ACADEMIC EMPHASIS

| Sr. No | University Classes | | |
|---|---|---|---|
| | Class | Count | % |
| 1 | ARTS | 114 | 47.89 |
| 2 | SCIENCE | 93 | 39.07 |
| 3 | COMMERCE | 24 | 10.08 |
| 4 | ENGINEERING | 92 | 38.65 |
| 5 | MANAGEMENT | 81 | 34.03 |
| 6 | EDUCATION | 28 | 11.76 |
| 7 | MEDICAL | 28 | 11.76 |

VIII. FUTURE DIRECTIONS

Recommending a university to a user requires our system to be aware of both user and university. Our work until has been concentrated in knowing about universities. We have made different classes of the universities based on their location, control, no of students, expenses required and many others which can be seen in the figure No 5. For our experimental uses we used existing dataset of North American universities. For our work we are planning to use only Indian universities. So the work of crawling information about Indian university is going on. Now our task is to get the profile information from social networking websites where the user is registered as a valid user. Now from this data obtained we will recommend universities from different classes to users.

IX. CONCLUSION

In this paper, we have presented an approach to automatically extract user and university information. We also refine and integrate this information to form our dataset. Automatic retrieval of user profiles relies on the social profile of the user on various social networking websites. We also processed this dataset in the Weka Tool to mine knowledge out if it. This helps us find out the user interest from the dataset.